\documentclass[12pt]{article}
\usepackage{epsf}
\title{Microscopic 
coloured quark-dynamics in the soft non-perturbative regime -- Description of hadron 
formation in relativistic S+Au collisions at CERN}

\author{S.~Scherer\dag\footnote{To whom correspondence should be addressed (scherer@th.physik.uni-frankfurt.de)}, 
M.~Hofmann, 
M.~Bleicher,
L.~Neise, 
H.~St\"ocker\dag,
W.~Greiner\dag} 

\date{4. June 2001}
\begin{document}
\maketitle 

\begin{indent}
\dag\ Institut f\"ur Theoretische Physik, J.W.Goethe-Universit\"at, Robert-Mayer-Str. 8-10, D-60054 Frankfurt am Main, Germany
\end{indent}

\begin{abstract}
The quark-molecular-dynamics model is used to study microscopically 
the dynamics of the coloured quark phase and the subsequent hadron formation 
in relativistic S+Au collisions at the CERN-SPS. Particle spectra and hadron 
ratios are compared to both data and the results of hadronic transport 
calculations. The non-equilibrium dynamics of hadronization and the loss 
of correlation among quarks are studied.
\end{abstract}

%%%%%%%%%%%%%%%%%%%%%%%%%%%%%%%%%%%%%%%%%%%%%%%%%%%%%%%%%%%%
%
% Text of paper follows
%
%%%%%%%%%%%%%%%%%%%%%%%%%%%%%%%%%%%%%%%%%%%%%%%%%%%%%%%%%%%%

\section{Introduction} 

The relativistic heavy ion collider (RHIC) at
Brook\-ha\-ven National Laboratory and the running Pb+Pb program at the CERN-SPS
aim to explore the phase diagram of hot and dense matter near the quark gluon
plasma (QGP) phase transition \cite{Bass:1998vz}. 
The QGP is a hypothetical state in which the individual hadrons dissolve into a
gas of free (or almost free) quarks and gluons in strongly compressed
and heated matter. The energy- and baryon densities achievable in heavy ion collisions
sensitively depend on the extend to which the nuclei are stopped during
penetration; they also depend on mass number and bombarding energy.

However, a detailed understanding of the quark dynamics and the subsequent
parton-hadron phase transition is still missing. First exploratory attempts have been
made using hydrodynamical simulations, like one-fluid hydro \cite{Rischke:1996nq} 
and three-fluid hydrodynamics \cite{Brachmann:1997bq,Brachmann:1999xt}. 
It has been found that the results of these investigations are strongly 
influenced by the freeze-out definition and the non-equilibrium features 
encountered in heavy ion reactions \cite{Brachmann:1997bq}. 

Very recently it has been tried to circumvent the freeze-out problem by
coupling the hydrodynamical simulation to the semi-hadronic 
transport model UrQMD~\cite{Bass:1999tu}. A detailed microscopic simulation 
of the quark phase and of the hadronization, however, is still missing. 

Attempts to simulate the dynamics of the hot and dense region  
within the VNI model~\cite{Bass:1999pv} suffer from
the non-perturbative origin of the quark-quark interaction at SPS and RHIC
energies -- this has already been pointed out by M\"uller and Harris, 
who stressed that it is important to realize that the QGP is a strongly interacting
gas of quarks and gluons at low momentum transfer $q^2 < q^2_{\rm CSB}$ (CSB
= Chiral symmetry breaking)~\cite{Harris:1996zx}. 
Even at RHIC energies leading order (and next-to leading order) perturbative QCD (pQCD) 
seems to be inapplicable to understand the heated and compressed system encountered as 
shown in \cite{Bass:1999bu,Fries:1999jj}. 

In this paper we present the first microscopic model calculation of the 
quark dynamics in the non-perturbative regime dubbed the Quark Molecular Dynamics 
Model (qMD)~\cite{Hofmann:1999jx,Hofmann:1999jy}. The model is applied to the dynamics 
of the expanding fireball in relativistic S+Au collisions at CERN.

\section{Discussion of the model}

The qMD model \cite{Hofmann:1999jx,Hofmann:1999jy} is a semi-classical model 
which mimics the properties of non-Abelian QCD by the means of a two-body colour
potential between quarks and antiquarks. In addition, a dynamical hadronization criterion is 
defined which allows for the consecutive migration from quark to hadronic degrees 
of freedom. A very similar model has been used before to study the spontaneous 
generation of density perturbations in the early universe triggered by the 
hadronization of the primordial quark--gluon plasma \cite{Crawford:1982yz}.  

The coloured and flavoured quarks (and antiquarks) are treated as classical particles
interacting via a Cornell potential with colour matrices \cite{Isgur:1985bm}. 
This interaction provides an effective description of the 
non-perturbative, soft gluonic part of QCD. 

The model does not treat the explicit dynamics and fragmentation of
hard gluons. In any application of the model the validity
of this approximation has to be checked.

Using these assumptions, the model Hamiltonian reads
\[
{\mathcal{H}} =
\sum_{i=1}^N\sqrt{p_i^2+m_i^2}+\frac{1}{2}\sum_{i,j}C_{ij}
V(\left\vert\mathbf{r}_i-\mathbf{r}_j\right\vert)
\]
where $N$ counts the number of particles in the system. The particles
are quarks or antiquarks carrying a fixed colour charge -- red, green, blue 
for quarks, anti-red- anti-green and anti-blue for antiquarks. Four quark 
flavours ($u,d,s,c$) with current masses $m_u=m_d=10$~MeV, $m_s=150$~MeV and $m_c=1.5$~MeV
are considered. In order to allow for a later mapping of quark clusters to hadronic
states, the particles carry spin and isospin quantum numbers.
The confining properties of the potential $V(r)$ are ensured 
by a linear increase at large distances~$r$. 
At short distances, the strong coupling constant
$\alpha_s$ becomes small, yielding a Coulomb-type behaviour as in QED.
For this colour Coulomb potential plus the confining part we use
the well known Cornell-potential \cite{Eichten:1975af}
\[
V(r) = -\frac{3}{4}\frac{\alpha_s}{r}+\kappa\,r\;,
\]
which has successfully been applied to meson spectroscopy.
For infinite quark masses this inter-quark potential
has also been found in lattice calculations over a wide
range of quark distances \cite{Born:1994cq}.
For small quark masses, retardation and
chromomagnetic effects should be included.
This is neglected in the present work. However, the linear
behaviour at large distances seems to be supported by the success of
the string model even for zero quark masses
\cite{Andersson:1983ia}. 

By fixing the colour of each quark or antiquark, we restrict ourselves to 
the approximation that there are no interactions changing the colour of quarks.
The use of this approximation -- which breaks, of course,  the full colour 
gauge symmetry -- is motivated by the physical picture behind the maximal Abelian 
gauge of QCD~\cite{Ezawa:1982bf} which gives, at the same time, a justification 
for the use of the linearly increasing potential.
In this gauge, there remains a residual gauge freedom corresponding to the two 
diagonal gluons. The non diagonal gluons become massive~\cite{Polikarpov:1997wd},
contributing only over very short ranges in the order 
0.2 -- 0.3 fm~\cite{Amemiya:1999yk,Amemiya:1999zf}. Simultaneously, the colour fields 
of the diagonal gluons are quenched to flux tubes via the dual Meissner 
effect~\cite{'tHooft:1981ht}. This mechanism then yields the linear increase of 
the colour potential with distance. The remaining colour interaction contains the 
diagonal gluons which carry no colour and therefore do not change colour.

\begin{table}
\caption{Colour matrix elements of the 36 different
elementary colour combinations of the quarks. The matrix
elements can be obtained from the scalar products of the 
corresponding weight vectors}
\vspace*{0.4cm}

\renewcommand{\arraystretch}{1.2}
\renewcommand{\arraycolsep}{5mm}
\begin{tabular}{l|cccccc}
\hline
$C_{\rm c}^{\alpha\beta}$&$R$&$G$&$B$&$\overline{B}$&$\overline{G}$&$\overline{R}$\\
\hline
$R$&$-1$&$+\frac{1}{2}$&$+\frac{1}{2}$&$-\frac{1}{2}$&$-\frac{1}{2}$&$+1$\\
$G$&$+\frac{1}{2}$&$-1$&$+\frac{1}{2}$&$-\frac{1}{2}$&$+1$&$-\frac{1}{2}$\\
$B$&$+\frac{1}{2}$&$+\frac{1}{2}$&$-1$&$+1$&$-\frac{1}{2}$&$-\frac{1}{2}$\\
$\overline{B}$&$-\frac{1}{2}$&$-\frac{1}{2}$&$+1$&$-1$&$+\frac{1}{2}$&$+\frac{1}{2}$\\
$\overline{G}$&$-\frac{1}{2}$&$+1$&$-\frac{1}{2}$&$+\frac{1}{2}$&$-1$&$+\frac{1}{2}$\\
$\overline{R}$&$+1$&$-\frac{1}{2}$&$-\frac{1}{2}$&$+\frac{1}{2}$&$+\frac{1}{2}$&$-1$\\
\hline
\end{tabular}
%%\parbox{4.5cm}{\hfill\epsfig{figure=bilder/root,width=4cm}}
\hfill
\parbox{5cm}{\epsfxsize=5cm \epsfbox{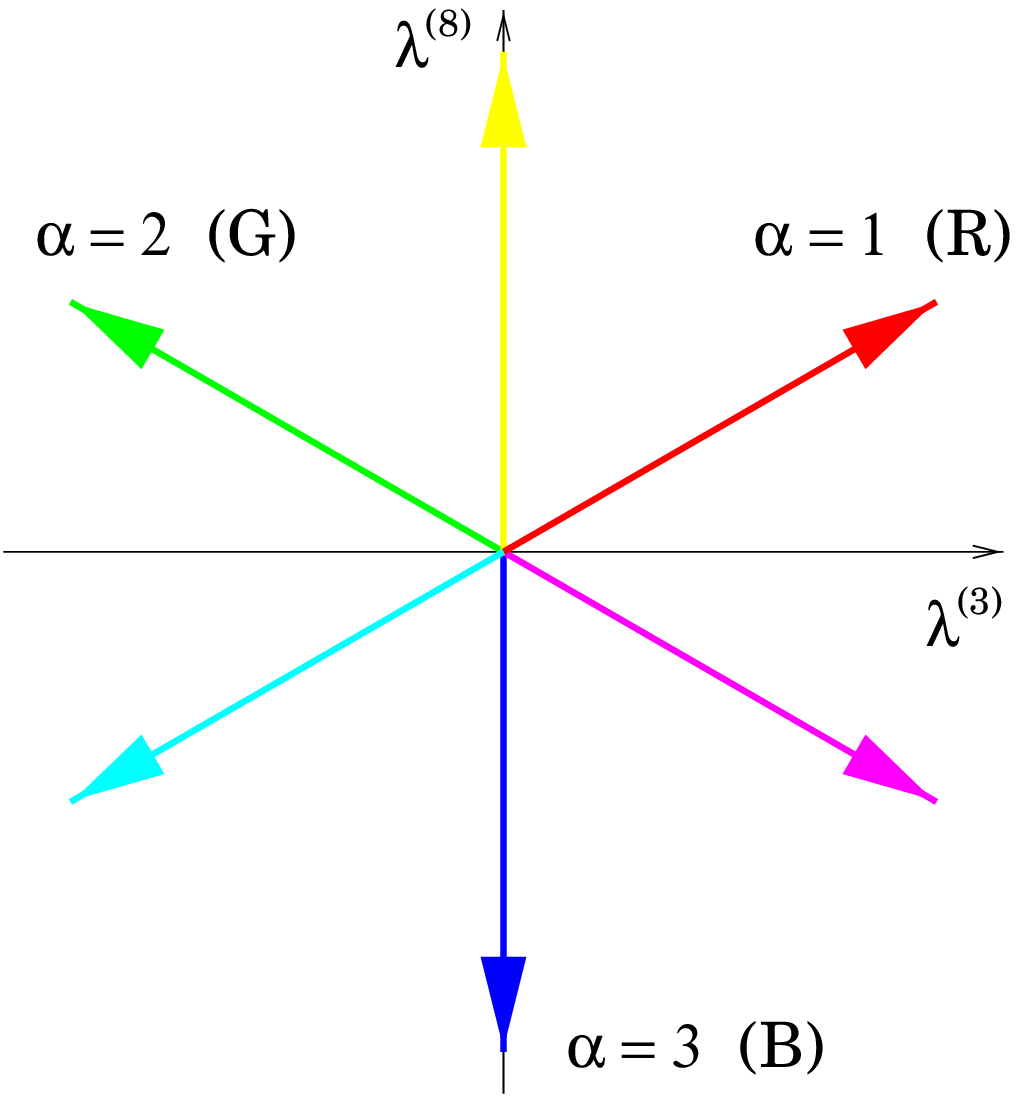}}
\label{tab1}
\end{table}

The colour matrix elements $C_{ij}$ regulate the sign and
relative strength of the interaction between two quarks/antiquarks,
respectively, depending on the colour combination of the pair. 
The matrix $C_{ij}$ in the short range colour 
interaction potential between quarks,
$V_{\rm colour} = - C_{ij}\frac{3}{4}\frac{\alpha}{r}$,
can be calculated from one-gluon exchange of the quark-gluon interaction 
of the QCD Lagrangian,
\[
{\mathcal{L}}_{\mathrm{int}} =
\frac{g}{2}\,\bar{\Psi}\lambda_a\gamma_\mu\Psi G^\mu_a\;.
\]
Using the standard fundamental representation of $SU(3)_{\rm colour}$ for the
quarks and the adjoint representation for the gluons,
\[
\vec{q}_R = \left(\begin{array}{c}
1\\ 0\\ 0
\end{array}\right),\quad
\vec{q}_G = \left(\begin{array}{c}
0\\ 1\\ 0
\end{array}\right),\quad
\vec{q}_B = \left(\begin{array}{c}
0\\ 0\\ 1
\end{array}\right),\quad
\]
\[
T^a = \frac{1}{2}\lambda^a,\quad a=1,\dots,8
\]
where $\lambda_a$ are the Gell-Mann matrices, and separating 
the quark wave function in the colour and Dirac parts, the interaction amplitude 
also splits in colour and Dirac parts:
\[
\mathcal{M}_{\alpha\alpha^\prime\beta\beta^\prime}
\sim
\bar\psi_1\gamma_\mu\psi_1D^{\mu\nu}(q)\psi_2\gamma_\nu\psi_2\,
\vec{q}_{\alpha^\prime}^{\;\dagger}\lambda^a\vec{q}_{\alpha}\delta_{ab}\vec{q}_{\beta^\prime}^{\;\dagger}\lambda^b\vec{q}_{\beta}\;.
\]
Here, $\mathcal{D}^{\mu\nu}_{ab}(q) = D^{\mu\nu}(q)\delta_{ab}$ 
is the gluon propagator, and $\alpha$ and $\beta$ represent the colour charges of the 
incoming quarks, $\alpha^\prime$ and $\beta^\prime$ of the 
outgoing quarks in the one-gluon exchange diagram.
Collecting the colour parts in a colour factor
\[
C^{\rm c}_{\alpha\alpha^\prime\beta\beta^\prime} =
\frac{3}{4}\sum_{a=1}^8\vec{q}_{\alpha^\prime}^{\;\dagger}\lambda^a\vec{q}_\alpha\,\vec{q}_{\beta^\prime}^{\;\dagger}\lambda^a\vec{q}_\beta
=
\frac{3}{4}\sum_{a=1}^8(\lambda^a)_{\alpha\alpha^\prime}\,(\lambda^a)_{\beta\beta^\prime}\;,
\]
one can calculate the net amplitude by summing over all possible
combinations of in- and outgoing colours. In the maximally Abelian gauge,
only the the diagonal gluons, described by the commutating Gell-Mann matrices 
$\lambda_3$ and $\lambda_8$ from the Cartan subalgebra of $SU(3)_{\rm colour}$, 
contribute over larger distances. In this approximation
the total colour matrix for quark-quark interactions then is given by
\[
C^{\rm c}_{\alpha\beta} 
= \frac{3}{4}\sum_{a=3,8}(\lambda^a)_{\alpha\alpha}(\lambda^a)_{\beta\beta}
= \vec{w}_{\alpha}^{\sf T}\vec{w}_{\beta}\;,
\]
where
\[
\vec{w}_{\alpha} = \frac{\sqrt{3}}{2}\left(\begin{array}{c}
(\lambda^3)_{\alpha\alpha}\\ (\lambda^8)_{\alpha\alpha}\\ 
\end{array}\right),\quad \alpha =  1,2,3\;(R, G, B)
\]
are the normalised weight vectors corresponding to the three quark colours
in $(\lambda^3,\lambda^8)$ space.
Imposing a factor $-1$ at each antiquark vertex in colour space 
yields the colour matrix elements for the different colour combinations 
as collected in table \ref{tab1}. They can easily be read off as the
scalar products of the weight vectors corresponding to the three colours or
anticolours, respectively. Positive values indicate attractive, negative
repulsive interactions. 

The classical time evolution of the (anti)quark system as given by the
Hamiltonian yields colourless clusters of (a) three quarks with different 
colours or (b) a quark and an antiquark with opposite colours.  This is a 
consequence of the interplay between the attractive and repulsive interactions 
in the system which trigger this segregation. These quark clusters are mapped 
on hadronic resonance states according to energy-momentum conservation and the 
known spin and isospin quantum numbers of the quarks in the cluster. In our model, 
the resonance states do not interact any more with the system (there is no hadronic 
rescattering) - they just decay into stable hadrons.

It should be noted that the qMD model includes the possible creation of 
quark-antiquark pairs out of the vacuum. However, this is a process which 
is rare in systems with no large fluctuations in colour charge. We will
come back on this point in our application of the model to heavy ion 
collisions.

A detailed discussion of the model, including the clustering and hadronization
procedure, resulting thermal properties and the 
Equation of State, and analytic test cases are found 
in \cite{Hofmann:1999jx,Hofmann:1999jy}. 

There are, of course, several other approaches to the dynamical
description of the quark degrees of freedom in a hot and dense quark system.
In \cite{Rehberg:1998me,Rehberg:1999qs}, a simulation based on the Nambu-Jona-Lasinio
model has been presented, where quarks are propagated on 
classical trajectories while their effective masses are 
calculated self-consistently according to the NJL equations of motion. 
Hadron production is driven by $qq$ and $qh$ collisions. 
Unfortunately, this approach does not provide confinement 
and therefore is less suited for the investigation of heavy-ion
collisions. A fully gauge symmetric treatment of the colour degrees of freedom 
of the quarks including colour rotations has been presented in \cite{Maruyama:1999ec}. 
However, in this approach it is difficult to regain colourless hadrons out of
the quark system. 
It is also possible to treat the colour field as a dynamical background field
whose source are the quarks \cite{Traxler:1998bk}. This approach (while also fixing 
the quark colour) is less idealised than the one presented in this paper, but the 
numerics is still too time consuming to treat systems of
the size of fireballs from heavy ion collisions.

\section{Application to heavy ion collisions at the SPS}

We now apply the qMD model to study the expansion and hadronization of the fireball 
created in heavy ion collisions between S and Au nuclei at the SPS. 

The initial state of the quark dynamics is generated by a transport simulation
with hadronic and string degrees of freedom (in this paper the UrQMD model
\cite{Bass:1998ca} is used---other models can also be applied). The initial nuclei are
propagated within the transport simulation until a critical energy density is reached. 
In the investigated collisions, we assume the dissolution of the hadronic degrees of freedom 
at an energy density of 1 GeV/fm$^{-3}$ in line with recent lattice calculations~\cite{Okamoto:1999ca}. 
This energy density is reached at full overlap, i.\,e.\ 1.5 fm/c after the begin of the interaction. 
At this time all hadrons (mesons and baryons) are dissolved and all newly produced
q$\bar{\rm q}$ pairs from the decaying colour fields are used as
input of the quark Molecular Dynamics model.

In the study presented here, it is justified to omit the explicit dynamics and 
fragmentation of thermal gluons due to their small number. The number $N$ of 
gluons in the interaction region of volume $V$ and temperature $T$ can be estimated by: 
\[
N = V g \int \frac{{\rm d}^3 p}{(2\pi)^3} \frac{1}{{\rm exp}(|p|/T)-1} 
= V \frac{g}{\pi^2}\zeta(3)T^3\quad,
\]
with the degeneracy $g= 16$. Assuming Bjorken-like initial expansion and a
temperature of $T=200$~MeV we expect 88 thermal gluons in our simulation at the
moment of full overlap, carrying approximately 10\% of the entropy of the system. 
Due to the cooling of the system at expansion, this provides an estimate 
for an upper bound on the number of thermal gluons. Thus, the additional fragmentation 
of these gluons is not expected to change the calculated yields and ratios significantly 
and has been left out in the present calculations. 
The contribution of hard gluons is negligible at SPS energies.

The results of the qMD simulation of a S+Au collision 
are visualised in two MPEG movies. 
The first movie depicts the time evolution 
%   ^^^^^^^^^^^
%   s200au_2.mpg
%
of a collision in the plane of the beam axis (horizontal 
axis, $z$) and the impact parameter axis (vertical axis, $x$). 
The second movie shows the same event, 
%   ^^^^^^^^^^^
%   s200au_beam.mpg
%
but this time as seen by an observer on the beam axis 
(the horizontal axis in this movie is the impact parameter axis, $x$). 
The second movie shows nicely the wobbling motion of the 
quarks and antiquarks (coloured spheres) forming clusters 
of two or three (carrying colour and anti-colour or three 
different colours, respectively) which are then considered 
as hadrons (grey spheres).

\begin{figure}
\begin{center}
\epsfbox{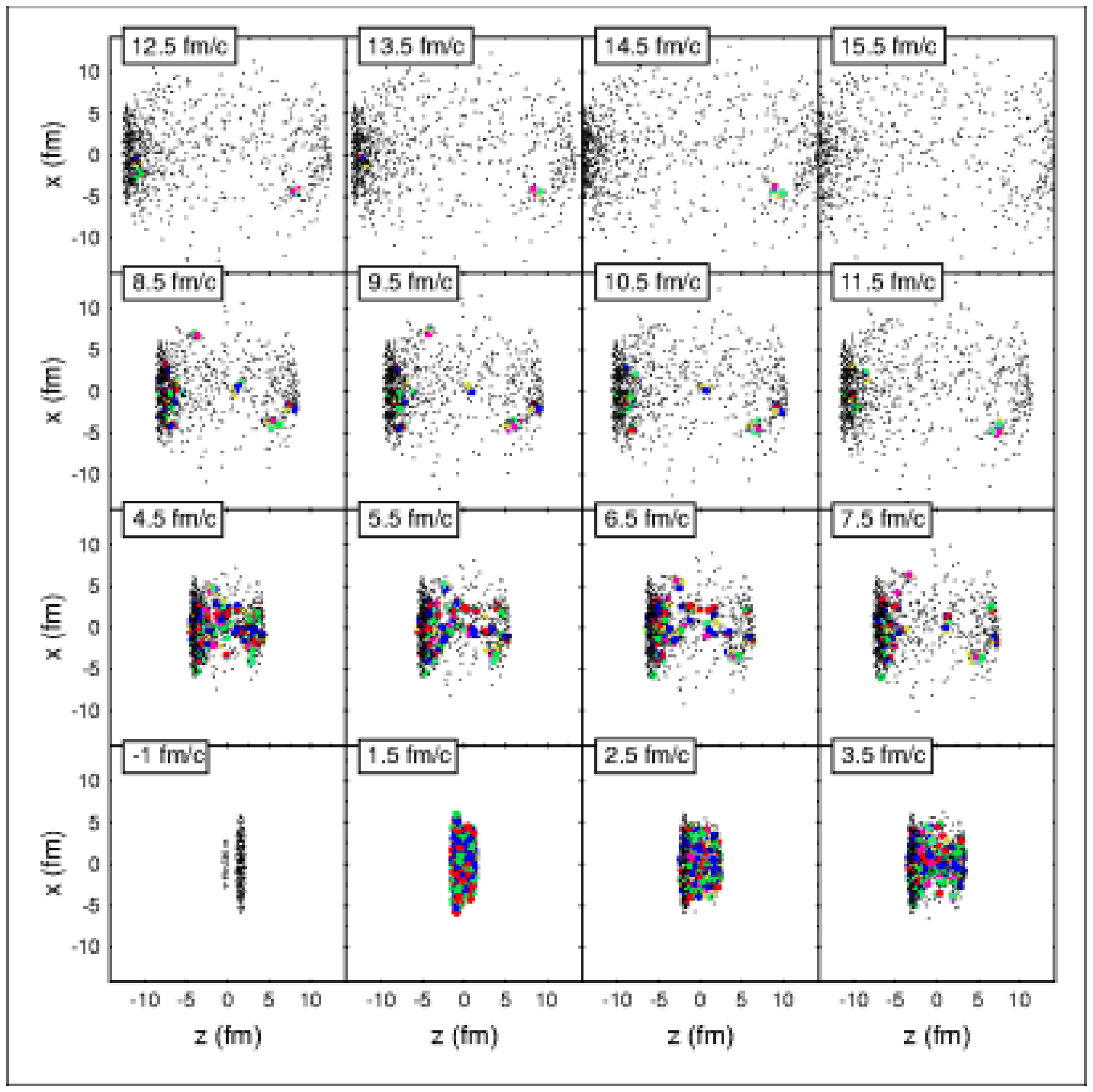}
\end{center}
\caption{Time evolution of quarks, antiquarks and hadrons in a central S+Au at SPS (200 GeV/$N$). 
$(z)$ is the beam axis and  $(x)$ the impact parameter direction.}
\label{figure1}
\end{figure}

Figure \ref{figure1} also shows the space-time evolution of the quarks and antiquarks (thick,
coloured dots) and hadrons (small white dots): The central zone of the collision 
is filled up with (anti)quarks that re-arrange themselves to colour neutral clusters. These
clusters are then decaying into physical hadron states. 

\begin{figure}
\begin{center}
\epsfxsize=\linewidth
\epsfbox{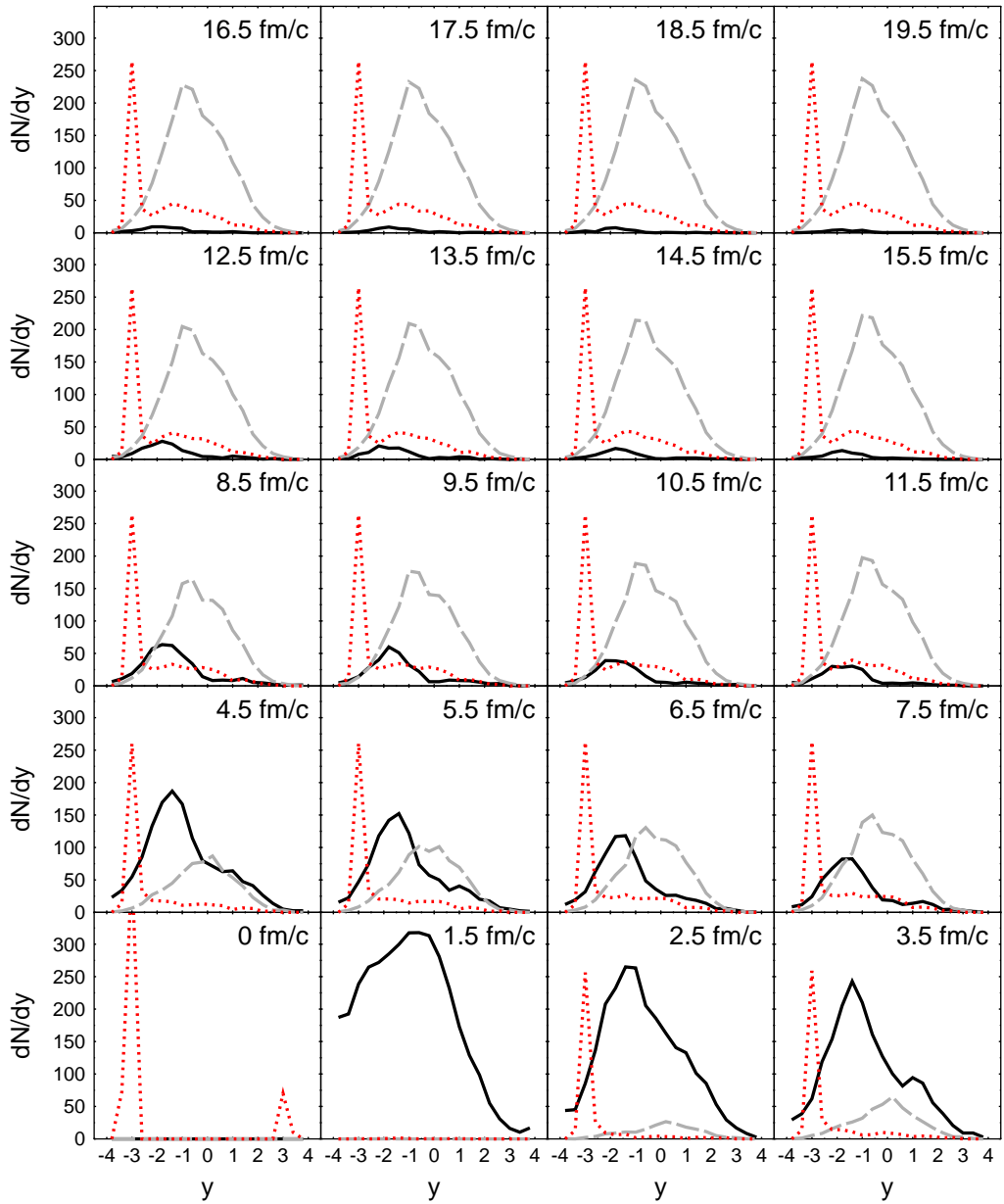}
\end{center}
\caption{Time evolution of the rapidity distributions of baryons (dotted lines), 
quarks and antiquarks (solid line) and mesons (dashed line) in a central S+Au 
collision at SPS (200 GeV/$N$).}
\label{figure2}
\end{figure}

Figure \ref{figure2} depicts the evolution of the rapidity densities of
baryons (dotted lines), mesons (dashed lines) and (anti)quarks (full
lines). The remnants of the gold nucleus re-hadronize immediately. The quarks 
at central rapidities start to form mesons. After 20 fm/c the quarks
have re-combined to hadrons and the collision ceases.

\begin{figure}
\begin{center}
\epsfxsize=\linewidth
\epsfbox{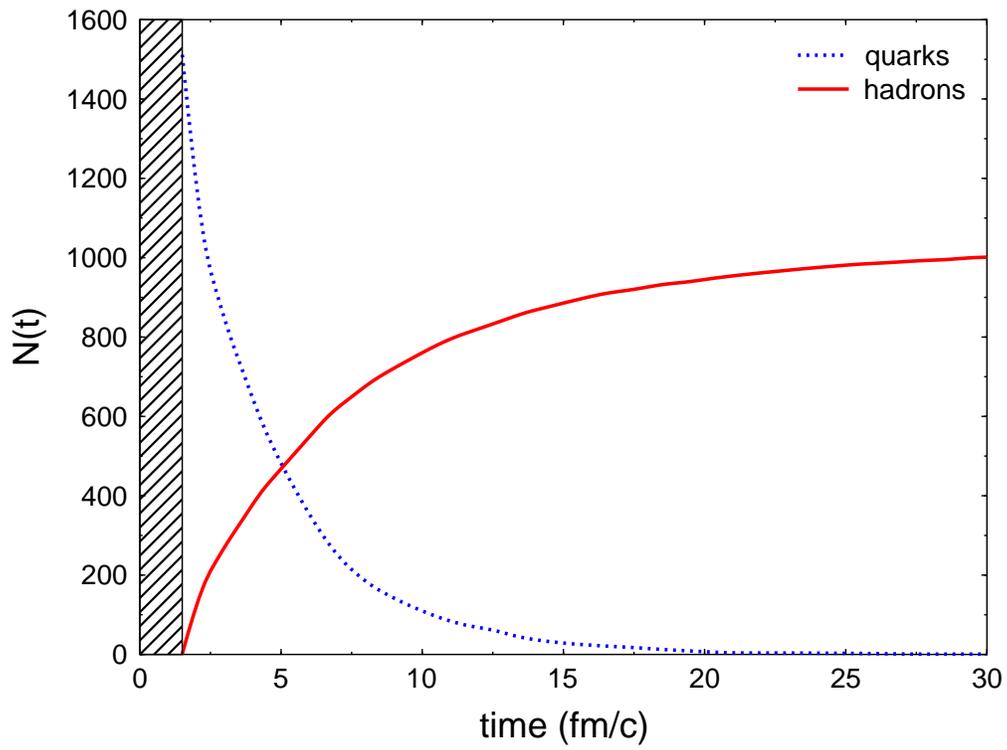}
\end{center}
\caption{Time evolution of the total number of (anti)quarks and hadrons in the system.}
\label{figure2a}
\end{figure}

The time evolution of the particle multiplicities (Figure~\ref{figure2a}) shows 
that the production of hadronic clusters starts immediately after the beginning of the collision. 
The hadron multiplicity rises steeply at $t=1.5$~fm/c, while the quark 
multiplicity goes down. The steep rise in the hadron abundance is due to
the rapid expansion of the initial hot and dense system.
At later times, the multiplicity curves saturate.
There could be two reasons for this flattening: either the back reaction $h h\to q\bar q$ 
becomes important due to the large abundance of pions, or the particle density becomes to low to
cause further collision-induced changes of the multiplicities. 
Since the remaining number of unbound quarks vanishes with the
saturation of the hadron multiplicity, we conclude in line with the 
findings in \cite{Rehberg:1998me} that the hadron numbers saturate due to the
breakup of the fireball rather than the approach to chemical equilibrium in the
$h h\leftrightarrow q\bar q$ channel.   

Note that the number of final hadrons is
larger then the number of 2- and 3-clusters which can be formed out of the quarks and antiquarks
in the initial state. This is caused by the decays of the hadronic clusters. 
As mentioned before, qMD includes the possible dynamical production of q$\bar{\rm q}$ pairs in 
overcritical colour fields. However, in the present model, hadronization at SPS 
is basically driven by quark rearrangement: On average, only 0.1 quark-antiquark pairs 
per event are produced by string breakup in a S+Au collision. The reason is that 
by dissolving the hadronic content of UrQMD one start with a 
configuration which is colour neutral on longer scales. Thus, colour fluctuations 
large enough to yield strong colour fields which will then drag quark-antiquark pairs out of 
the vacuum are never reached. Moreover, strong fields are dynamically screened by the 
moving quarks.

\begin{figure}
\begin{center}
\epsfxsize=\linewidth
\epsfbox{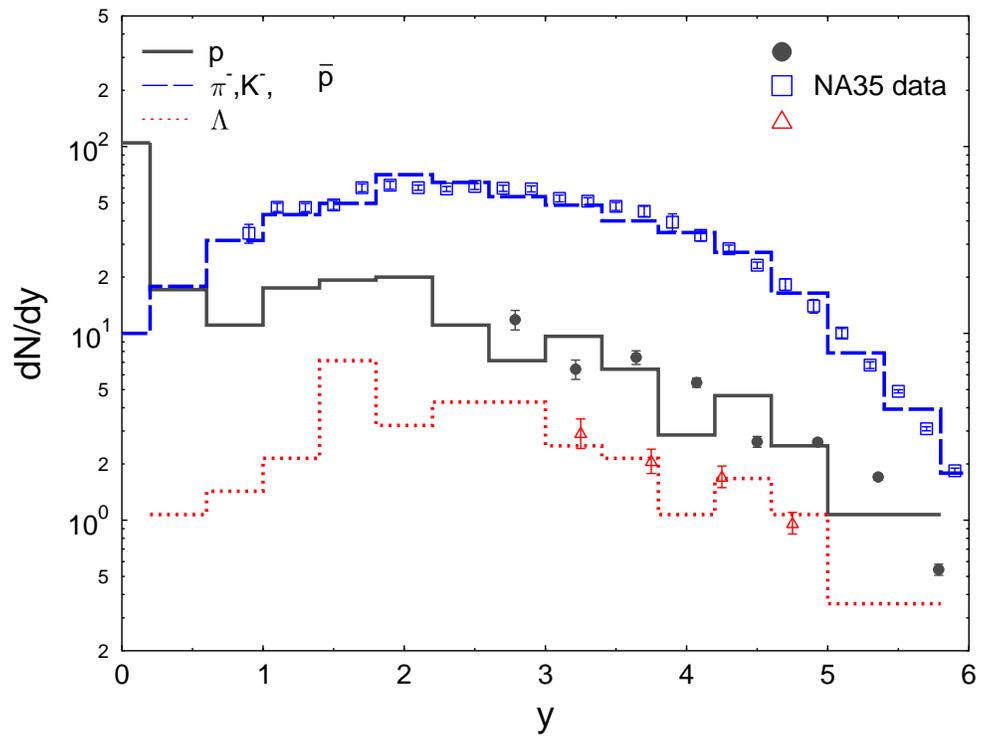}
\end{center}
\caption{Rapidity distributions of protons, $\Lambda$'s and $h^-$ in central S+Au collisions 
at SPS (200 GeV/$N$) compared to NA35 data~\cite{PBM}.}
\label{figure3}
\end{figure}

Let us now confront the qMD results to experimental data. Figure~\ref{figure3}
compares the calculated proton (full line), lambda (dotted line) and $h^-$
rapidity densities to NA35 data (see \cite{PBM} and references therein). 
A surprisingly good description of the measured data (symbols) over all rapidities 
by the non-equilibrium quark dynamics is found.

\begin{figure}
\begin{center}
\epsfxsize=\linewidth
\epsfbox{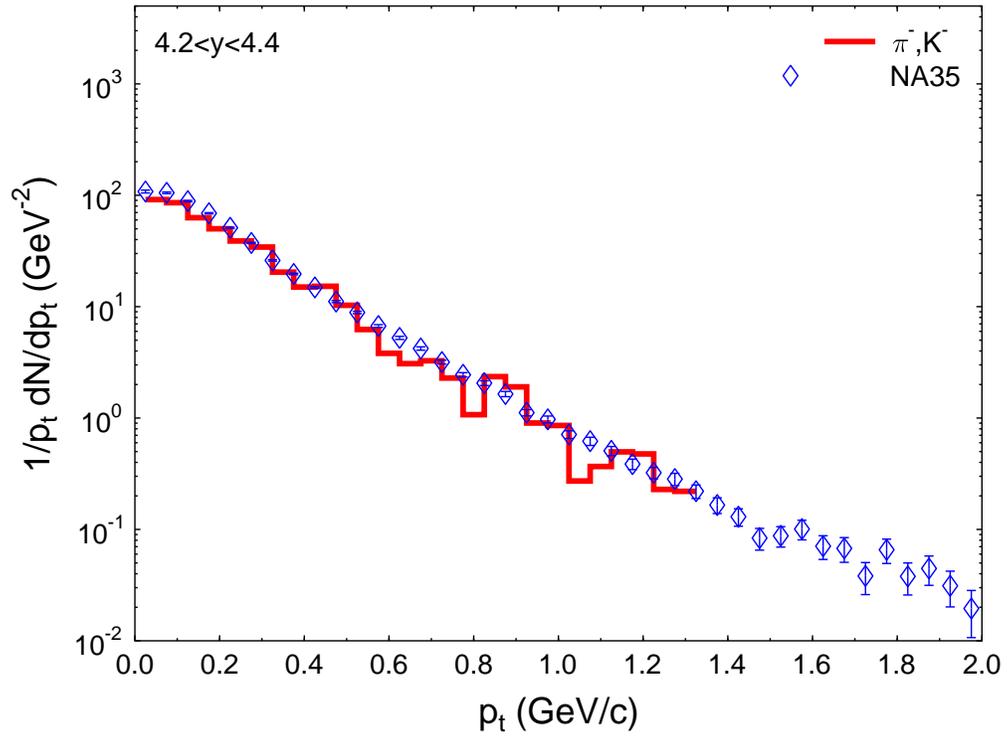}
\end{center}
\caption{Transverse mass spectrum of negative charged hadrons in central S+Au collisions 
at SPS (200 GeV/$N$) compared to NA35 data~\cite{PBM}.}
\label{figure4}
\end{figure}

Figure \ref{figure4} explores the the transverse expansion of the hadronizing
quark system. The transverse momentum spectrum of negatively charged
particles is compared to NA35 data. Overall good agreement is found. 

\begin{figure}
\begin{center}
\epsfxsize=\linewidth
\epsfbox{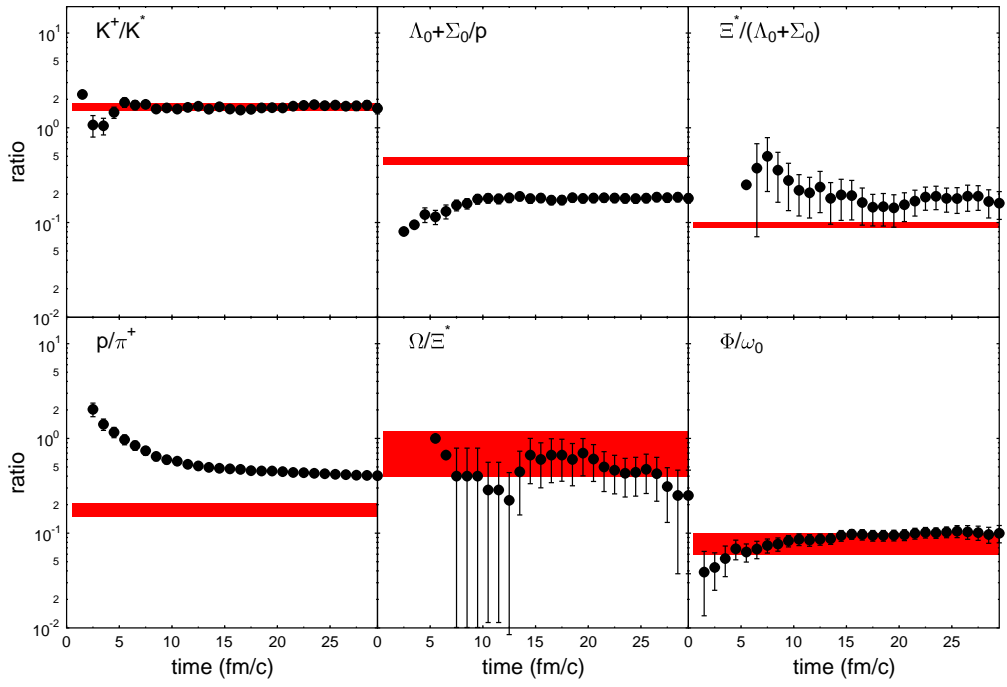}
\end{center}
\caption{Time evolution of the particle ratios in central S+Au collisions at SPS (200 GeV/$N$).
The shaded areas indicate data taken from~\cite{PBM}.}
\label{figure5}
\end{figure}

\begin{figure}
\begin{center}
\epsfxsize=\linewidth
\epsfbox{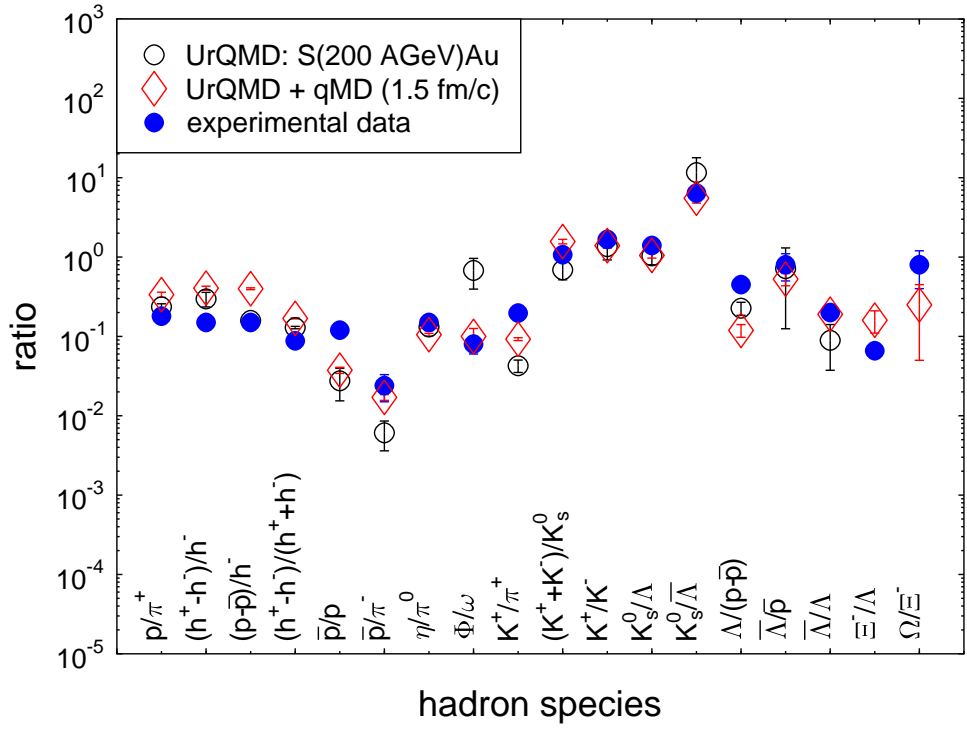}
\end{center}
\caption{Particle ratios in central S+Au collisions at SPS (200 GeV/$N$).
Experimental data are taken from \cite{PBM}.}
\label{figure6}
\end{figure}

Finally we want to investigate the hadro-chemical evolution and the behaviour 
of hadron ratios in the qMD model. Figure~\ref{figure5} displays the time 
development of several hadron ratios in the course of the hadronization 
(shaded areas denote the measured ratios). The particle ratios stay nearly 
constant during the parton-hadron conversion. This may explain the success 
of statistical model analysis \cite{Rafelski:1999qm}. However, baryons seem 
to be produced in the later stage of the hadronization. The discrepancies between 
the ratios of $\mathrm{p}/\pi^+$ and $(\Lambda_0+\Sigma_0)/\mathrm{p}$ obtained
from our approach and the measured data can be understood from the
missing hadronic rescattering in our approach. This can be seen by comparing
the measured hadron ratios in figure \ref{figure6} not only with the qMD calculation (diamonds) 
but also to ratios obtained with full UrQMD simulations (open circles). 

While the proton to antiproton ratio remains unaltered as compared to the UrQMD calculation 
alone (and nearly one order of magnitude lower than the experimental value), ratios involving
protons or antiprotons alone or the net proton number are by factors 2--5 higher
in qMD than UrQMD, yielding a better fit to data for the $\bar \mathrm{p}/ \pi^-$ ratio,
but an increasing overestimation for the $\mathrm{p}/ \pi^+$ and the $(\mathrm{p} - \bar \mathrm{p})/\mathrm{h}^-$
ratio. Simultaneously, the ratio of $\Lambda/(\mathrm{p} - \bar \mathrm{p})$ drops against 
the value from UrQMD. Both trends can be understood as consequences of hadronic rescattering in the UrQMD
model. Hadronic rescattering lowers the number of antiprotons in the final state, and, 
by inelastic baryon-baryon and antikaon-baryon collisions, yields a systematic population 
of hyperons at the expense of protons. These channels are not implemented in the present 
version of our qMD model. In qMD, delta resonances for example (which are abundant 
around the first 10 fm/c of the collision) cannot create hyperons and 
kaons by inelastic collisions, but eventually yield always nucleons and pions.
As this simplification of the combined UrQMD+qMD approach with respect to a full UrQMD 
treatment shifts the calculated ratios away from the known experimental values, this can 
be seen as a clear sign for the importance of hadronic rescattering.

As the hadron ratios in the combined approach show differences due to missing rescattering 
and the transverse mass spectra and rapidity distribution of figures \ref{figure4} and \ref{figure3} 
can also be reproduced by the UrQMD model~\cite{Bass:1998ca,Bleicher:1998wv}, one may
ask to which extent this combination of models may be useful or necessary.

The qMD approach can provide us with detailed information about the dynamics of 
the quark system and the parton-hadron conversion. Whenever a new hadron is formed, 
the correlation between the quarks clustering to build this hadron can be studied. 
The mean value of the path length these quarks have travelled within the quark phase 
from their points of origin until the clustering point and the distance 
in space between the origins of the involved quarks can be studied.

\begin{figure}
\begin{center}
\epsfxsize=\linewidth
\epsfbox{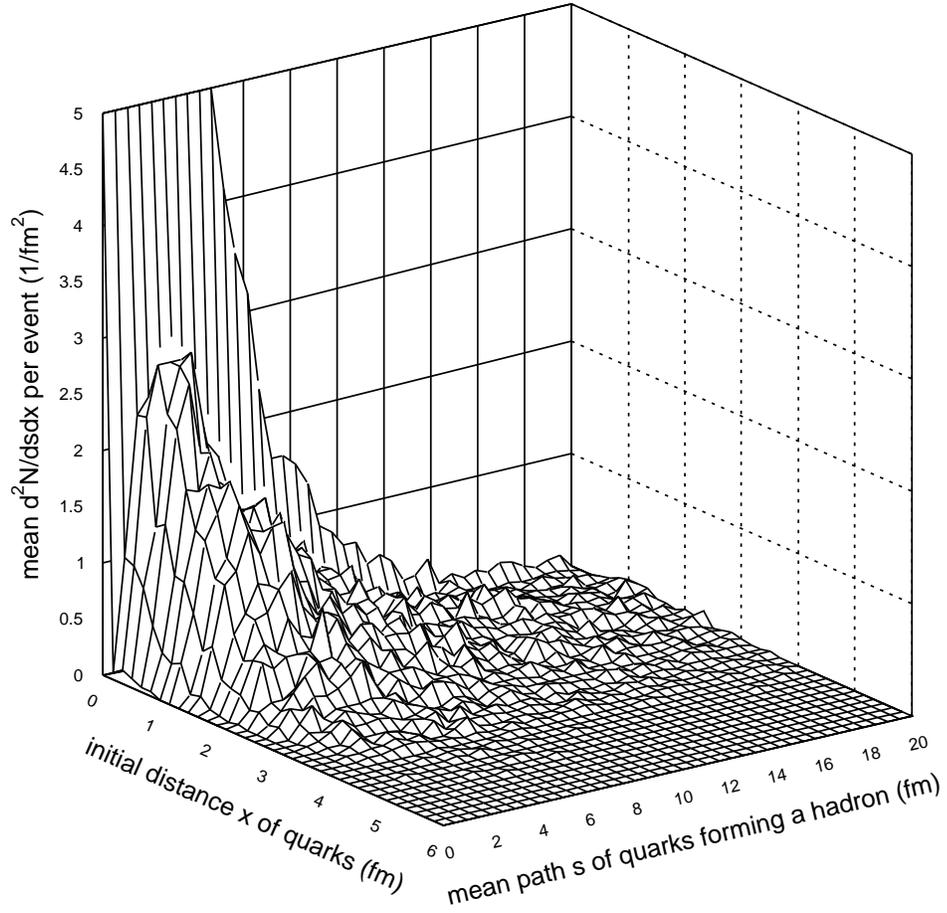}
\end{center}
\caption{Hadronization in S+Au collisions at SPS (200 GeV/$N$): Number density distribution of 
mean diffusion path and initial distance of quarks forming a hadron within qMD.}
\label{figure7}
\end{figure}

\begin{figure}
\begin{center}
\epsfxsize=\linewidth
\epsfbox{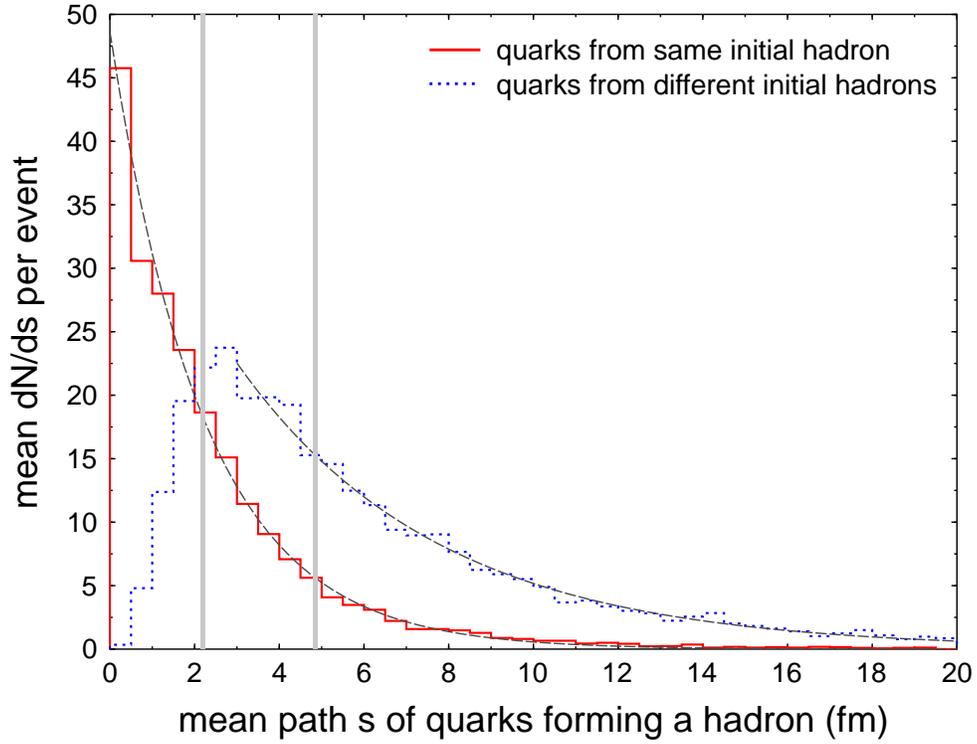}
\end{center}
\caption{Hadronization in S+Au collisions at SPS (200 GeV/$N$): Number density distribution 
of mean diffusion path of quarks forming a hadron from the same initial hadron (solid line) 
and from different initial hadrons (dashed line) within qMD. Fitting the decay profiles yields
diffusion lengths of $2.2\,\mathrm{fm}$ and $4.8\,\mathrm{fm}$, respectively
}
\label{figure7a}
\end{figure}

Figure \ref{figure7} shows the combined distribution of the mean path length 
travelled by the quarks and the original distance between the quarks 
forming a new hadron in S+Au collisions at SPS energies of $200\,\mathrm{GeV}/N$. 
Quarks stemming from the same initial hadron correlation, propagating coherently 
and reclustering again to form the same hadron build up a large background at zero 
initial distance. The essence of this distribution is contained in figure \ref{figure7a}
which shows the number distribution for the mean path travelled by quarks
forming a hadron (a) from the same initial hadron (solid line) and (b) from
different initial hadrons (dotted line). 

One sees that the reclustering in (a) is quite quick: in this case 
a hadron is decomposed into two or three quarks, these quarks propagate a short distance
of about $2.2\,\mathrm{fm}$ (``diffusion length'' -- marked by the grey box) and rehadronize again. 
The hadron formation follows an exponential decay of the quark cluster to hadrons with a decay length equal
to the diffusion length of $2.2\,\mathrm{fm}$.
 
On the other hand, the rearrangement of quarks to form new, different hadrons 
happens on a length scale of about $3\, \mathrm{fm}$. Following this rearrangement,
the clusters decay exponentially with a decay length of $4.8\, \mathrm{fm}$.

A measure of the relative mixing within the quark system -- and also for thermalization which
means homogenisation of the population of phase space and complete loss of correlations -- 
are the relative number of hadrons formed by quarks stemming from the same hadron correlation 
(rehadronization -- these hadrons are dubbed ``direct hadrons'') versus hadrons formed by quarks 
stemming from different hadrons (``mixed hadrons''). The ratio $r$ of mixed hadrons to the total
number of hadrons formed from the quark system is $r = 0.574 \pm 0.008$ for the S+Au collision at
$200\,\mathrm{GeV}/N$. The Error is from statistics. A value of $r=1$ would indicate complete rearrangement 
of quarks and thus complete loss of correlations in the quark system. 
Considering the presumed transition to the quark-gluon plasma in Pb+Pb
collisions at $160\,\mathrm{GeV}/N$, one would expect a much larger value of $r$.

We see that while global observables do not show large differences 
between hadronic calculations from UrQMD alone and our combined description
using quark molecular dynamics, the easy rearrangement of quarks does indeed
play a role and may hint to state of deconfined matter at the CERN SPS. 
Note that this rearrangement may be observed with the help of
balance functions \cite{Bass:2000az}. Such an analysis would, however, require 
better statistics.

\section{Conclusion}

In conclusion, we have used the Quark Molecular Dynamics Model (qMD) and
investigated the non-equilibrium quark dynamics at SPS energies. The qMD has
been coupled to the UrQMD model to generate the initial quark distributions. 
Good agreement with S+Au data at 200 GeV/$N$ is found. This includes
rapidity spectra, transverse momentum spectra and hadron ratios of mesons. 
While the missing rescattering in the qMD distorts the relative numbers of 
protons and hyperons, the coupled approach allows a detailed look at the dynamics of 
the quark system and the parton-hadron conversion. Our investigation shows that 
the parton-hadron phase transition in this model is mainly driven by quark re-arrangement. 
Quark-antiquark pair production is a rare process which needs strong colour fields 
which are, however, dynamically screened by the moving quarks.

\section*{Acknowledgements}

This work is in part supported by the BMBF, GSI, DFG and Graduiertenkolleg
``Theoretische und experimentelle Schwerionenphysik''.

%%%%%%%%%%%%%%%%%%%%%%%%%%%%%%%%%%%%%%%%%%%%%%%%%%%%%%%%%%%%
%
% literature:
%
%%%%%%%%%%%%%%%%%%%%%%%%%%%%%%%%%%%%%%%%%%%%%%%%%%%%%%%%%%%%

\end{document}